\documentclass[fleqn,usenatbib]{mnras}

\usepackage{newtxtext,newtxmath}
\usepackage[T1]{fontenc}

\DeclareRobustCommand{\VAN}[3]{#2}
\let\VANthebibliography\thebibliography
\def\thebibliography{\DeclareRobustCommand{\VAN}[3]{##3}\VANthebibliography}

\usepackage{lipsum}
\usepackage{graphicx}	% Including figure files
\usepackage{amsmath}	% Advanced maths commands
\newcommand{\new}{\color{black}}
\newcommand{\ie}{\textit{i}.\textit{e}.}
\title[Diffusion in WD Mixtures]{Accurate Diffusion Coefficients for Dense White Dwarf  Plasma Mixtures}

\author[M. E. Caplan, E. B. Bauer,  \& I. F. Freeman]{
M. E. Caplan,$^{1}$\thanks{E-mail: mecapl1@ilstu.edu}
Evan B. Bauer,$^{2}$
I. F. Freeman$^{1}$
\\
$^{1}$Department of Physics, Illinois State University, Normal, IL 67190, USA\\
$^{2}${Center for Astrophysics $\vert$ Harvard \& Smithsonian, 60 Garden St, Cambridge, MA 02138, USA}
}

\date{Accepted XXX. Received YYY; in original form ZZZ}

\pubyear{2021}

\begin{document}
\label{firstpage}
\pagerange{\pageref{firstpage}--\pageref{lastpage}}
\maketitle

\begin{abstract}

Diffusion coefficients are essential microphysics input for modeling white dwarf evolution, as they impact phase separation at crystallization and sedimentary heat sources. Present schemes for computing diffusion coefficients are accurate at weak coupling {\new ($\Gamma \ll 1$)}, but they have errors as large as a factor of two in the strongly coupled liquid regime {\new ($1 \lesssim \Gamma \lesssim 200$)}. With modern molecular dynamics codes it is possible to accurately determine diffusion coefficients in select systems with percent-level precision. In this work, we develop a theoretically motivated law for diffusion {\new coefficients} which works across the wide range of parameters typical for white dwarf interiors. We perform molecular dynamics simulations of pure systems and two mixtures that respectively model a typical-mass C/O white dwarf and a higher-mass O/Ne white dwarf, and resolve diffusion coefficients for several trace neutron-rich nuclides. We fit the model to the pure systems and propose a physically motivated generalization for mixtures. We show that this model is accurate to roughly {\new 15\%} when compared to molecular dynamics for many individual elements under conditions typical of white dwarfs, and is straightforward to implement in stellar evolution codes. 

\end{abstract}


\section{Introduction}

In recent years, observers have resolved core crystallization in white dwarfs (WDs) at the population level, as the latent heat of crystallization delays cooling \citep{van1968crystallization,Winget_2009,Gaia2018,Tremblay2019}. Data Release 2 from the {\it Gaia} mission has likewise identified a population of massive WDs with an anomalous heat source (the `Q branch') which is now thought to be caused by the settling of neutron-rich nuclides toward the core \citep{cheng2019cooling,Bauer2020,Camisassa2021,Blouin2021}. Beyond heating, the increased core density from sedimentation may be relevant for delayed supernova ignition in merger remnants \citep{Caiazzo2021} and other WD modeling such as pulsations \citep{Camisassa2016, Chidester2021}. 

These observations have motivated a rapidly growing body of theoretical work that is needed to accurately model WD interiors and heat sources, including binary phase diagrams \citep{Blouin2020,Blouin2021a}, ternary phase diagrams as mechanisms for precipitation and distillation \citep{Blouin2021,Blouin2021b,caplan2020neon,Caplan2021iron}, and micro-physics input such as diffusion coefficients \citep{Bauer2020,caplan2021mnras}. The drift velocity for sedimentation (and in the case of precipitation, cluster sizes) scale with the diffusion coefficients, so these coefficients are therefore an important microphysics input for accurately modeling WD evolution. %The diffusion coefficients can also have important impacts on other aspects of WD modeling such as pulsations \citep{Camisassa2016, Chidester2021}.

In the dense liquid regime of WD interiors, molecular dynamics (MD) methods can calculate diffusion coefficients to high accuracy, but the expense of MD has so far limited these calculations to coefficients for idealized one-component systems or a few elements in specific mixtures \citep{Hansen1975,Hughto2010,Daligault2012b}. It has so far been unclear how to extend these calculations to the general case of mixtures of many elements across the parameter space representative of WD interiors. Current methods used in stellar evolution codes for calculating diffusion coefficients in mixtures are based on binary collision integrals \citep{Paquette1986,Stanton2016}, and \cite{Bauer2020} have recently shown that these methods are only accurate to within about a factor of two when comparing to the available MD modeling in the regime of liquid WD interiors. As a consequence, uncertainties in heating and abundances of neutron-rich nuclei may also be of this order.

{\new For these reasons we develop in this work a simple, theoretically motivated model to accurately compute diffusion coefficients in mixtures for the entire range of conditions that may occur in WDs. This model builds on the work of \cite{Bildsten2001}, \cite{Daligault2006}, and \cite{Bauer2020} and is straightforward and efficient to implement in stellar evolution codes}. We show that scaling with charge can yield coefficients {\new accurate to about 10\%} when comparing to MD calculations for eight different elements in two distinct mixtures.
{\new Our scaling law is therefore validated against MD for many elements in two independent plasma mixtures, suggesting that it is generally applicable to typical WD plasma mixtures.}

We begin in sec.~\ref{sec:stanton} by describing the current scheme for calculating diffusion coefficients in stellar evolution codes. 
In sec.~\ref{sec:md} we describe our MD formalism and simulations of pure systems as well as realistic C/O and O/Ne WDs, and show the disagreement with the method of sec.~\ref{sec:stanton}. In sec.~\ref{sec:model} we present our analytic model for diffusion and verify its accuracy in comparison to MD. We conclude and discuss astrophysical implications in sec.~\ref{sec:dis}.

\section{Stanton \& Murillo Coefficients}\label{sec:stanton}

The current state of the art for diffusion coefficients in {\new mixtures for} stellar evolution is represented by \cite{Paquette1986} and \cite{Stanton2016}, e.g.\ see implementation of diffusion in the stellar evolution codes MESA \citep{Paxton2015,Paxton2018}, STELUM \citep{Bedard2021}, and LPCODE \citep{Althaus2001}.%
{\new \footnote{There is an important exception in the specific case of $^{22}$Ne settling in C/O WD interiors, where recent results have generally adopted the \cite{Hughto2010} coefficient based on MD for $^{22}$Ne \citep{Camisassa2016,Bauer2020,Salaris2022}. }
}
These coefficients rely on fits to numerical collision integrals for binary interactions between ions with a screened Coulomb potential, which cannot fully account for many-body dynamics at strong coupling. Careful choices of the screening length to account for both electron and ion interactions only allow an approximate accounting for many-body interaction effects in the moderately to strongly coupled regime.
Here we provide an implementation of the \cite{Stanton2016} coefficients in the white dwarf plasma regime for comparison to our later MD results and fits.

We start by considering a plasma with ion spacing $a = (4 \pi n/3)^{-1/3}$ (ion number density $n$) and electron screening length $\lambda_e$. For a one-component plasma (OCP) with ion charge $eZ$, the dimensionless plasma coupling $\Gamma = e^2 Z^2 / a k_{\rm B}T$ is an effective temperature, and the plasma is solid (liquid) above (below) $\Gamma_{\rm crit} \approx 175$, with a weak dependence on $\kappa \equiv a/\lambda_e$ when $\kappa \lesssim 1$. The OCP is fully characterized by the two parameters $\Gamma$ and $\kappa$. Generalized to mixtures, each species of charge $Z_i$ has $\Gamma_i = e^2 Z_i^2 / a_i k_{\rm B}T$ where $a_i$ is defined from the average charge density of the mixture $n_{e} = \langle Z \rangle n$, such that $a_i = (3Z_i / 4 \pi n_{e})^{1/3}$. Averaging over all species gives multi-component plasma coupling $\Gamma_{\rm MCP} = e^2 \langle Z^{5/3} \rangle / a_e k_{\rm B}T$ with $a_e = (3 / 4 \pi n_{e})^{1/3}$.
{\new Weak plasma coupling is defined by $\Gamma \ll 1$, while the strongly-coupled liquid regime is $1 \lesssim \Gamma \lesssim 200$, and crystallization into the solid phase generally occurs for $\Gamma_{\rm MCP} \gtrsim 200$ in plasma mixtures.}

The diffusion coefficients can be expressed in terms of the \cite{Burgers1969} resistance coefficients, which are
\begin{equation}
K_{ij} = \frac{16}{3} n_i n_j \mu_{ij} \Omega_{ij}^{(1,1)}~,
\end{equation}
where $\Omega_{ij}^{(n,m)}$ are the binary collision integrals \citep{Chapman1970,Ferziger1972}, $n_i$ is the number density for ions of type $i$, and $\mu_{ij} = m_i m_j / (m_i + m_j) $ is the reduced mass for particles of type $i$ and $j$. We use the fits given in the appendix of \cite{Stanton2016} for the collision integrals $\Omega_{ij}^{(n,m)}$. These fits depend on an effective screening length $\lambda_{\rm eff}$, which accounts for both electron and ion screening terms. For the electrons, we use a relativistic screening length $\lambda_e^{-1} = 2 k_F \sqrt{\alpha/\pi}$  {\new typical for $\rho \gg 10^6 {\rm \, g/cm}^3$} (electron Fermi momentum $k_F = (3 \pi^2 \langle Z \rangle n)^{1/3}$, %ion number density $n$, 
fine-structure constant $\alpha$) to match the form assumed for our MD calculations. For the ion contribution to the screening, we follow the suggestion of \cite{Stanton2016} for mixtures:
\begin{equation}
\lambda_{\rm ions} = \left[\sum_i \frac{1}{\lambda_i^2} \left( \frac{1}{1 + 3 \Gamma_i} \right) \right]^{-1/2}~,
\end{equation}
where $\lambda_i = (k_{\rm B}T / 4 \pi Z_i^2 e^2 n_i)^{1/2}$ 
%\begin{equation}
%\lambda_i = \sqrt{\frac{k_{\rm B}T}{4 \pi Z_i^2 e^2 n_i}}
%\end{equation}
is the Debye length for each species $i$. The net effective screening length is then
$\lambda_{\rm eff} \equiv (\lambda_e^{-2} + \lambda_{\rm ions}^{-2})^{-1/2}$. Note that no analogous ion screening term ever appears directly in the evaluation of the MD because all ion-ion interactions are explicitly accounted for in the case of MD, whereas in the collision integral formalism the screening term is the only way to account for many-body effects in an otherwise strictly two-body calculation. This treatment based on screened binary interactions works well in the diffuse limit $\Gamma \ll 1$. As seen in later sections, with the right choice of screening length it also yields coefficients that are correct to within a factor of a few for $\Gamma > 1$, but fails to capture the effect of caging due to many-body interaction at $\Gamma \gg 1$.

The diffusion coefficient for species $j$ in a mixture is
\begin{equation}
D_j = \frac{n_j k_{\rm B} T}{\sum_i K_{ij}}~,
\end{equation}
and its dimensionless form is $D^*_j = D_j/\omega_p a^2$ with ion plasma frequency $\omega_p = ( 4 \pi e^2 \langle Z \rangle^2 n / \langle M \rangle )^{1/2}$ where $\langle M \rangle$ is the average ion mass. Note that our expression here neglects a second-order correction to the diffusion coefficient that accounts for thermal diffusion with higher order collision integrals, which is often written as $1/(1 - \Delta)$. This correction can be up to 20\% in the weakly-coupled limit, but is negligible for $\Gamma > 10$ \citep{Baalrud2014}.

We evaluate diffusion coefficients for $\Gamma = 1-300$ in WD plasma mixtures assuming a density of $\rho = 10^6 \, \rm g\,cm^{-3}$, and compare to MD in the next section.

\section{Molecular Dynamics}\label{sec:md}

Our MD formalism is the same as in our past work, see \cite{caplan2021mnras,caplan2020neon,Caplan2021iron}. To briefly review, nuclei in WDs are fully ionized and surrounded by a neutralizing background of degenerate electrons. Nuclei are treated as classical point particles of charge $eZ$ experiencing a Coulomb repulsion screened by electrons, 

\begin{equation}\label{eq.V}
    V_{ij}(r) = \frac{e^2 Z_i Z_j}{r_{ij}} e^{-r_{ij}/\lambda_e}~,
\end{equation}

\noindent with inter-particle separation $r_{ij}$ and screening length $\lambda_e$. %In WD cores we assume fully relativistic electrons with screening length $\lambda_e^{-1} = 2 k_F \sqrt{\alpha/\pi}$. 
In this work, we use the dimensionless screening parameter $\kappa = a/\lambda_e$; $0.2 \lesssim \kappa \lesssim 0.4$ are typical for WDs \citep{caplan2021mnras,Blouin2021a}. Simulations contain 65536 nuclei and the force is computed to the nearest periodic image for all pairs and evolved using velocity Verlet.

\begin{figure}
\centering  %left bottom top right)
\includegraphics[trim=0 120 180 0,clip,width=0.48\textwidth]{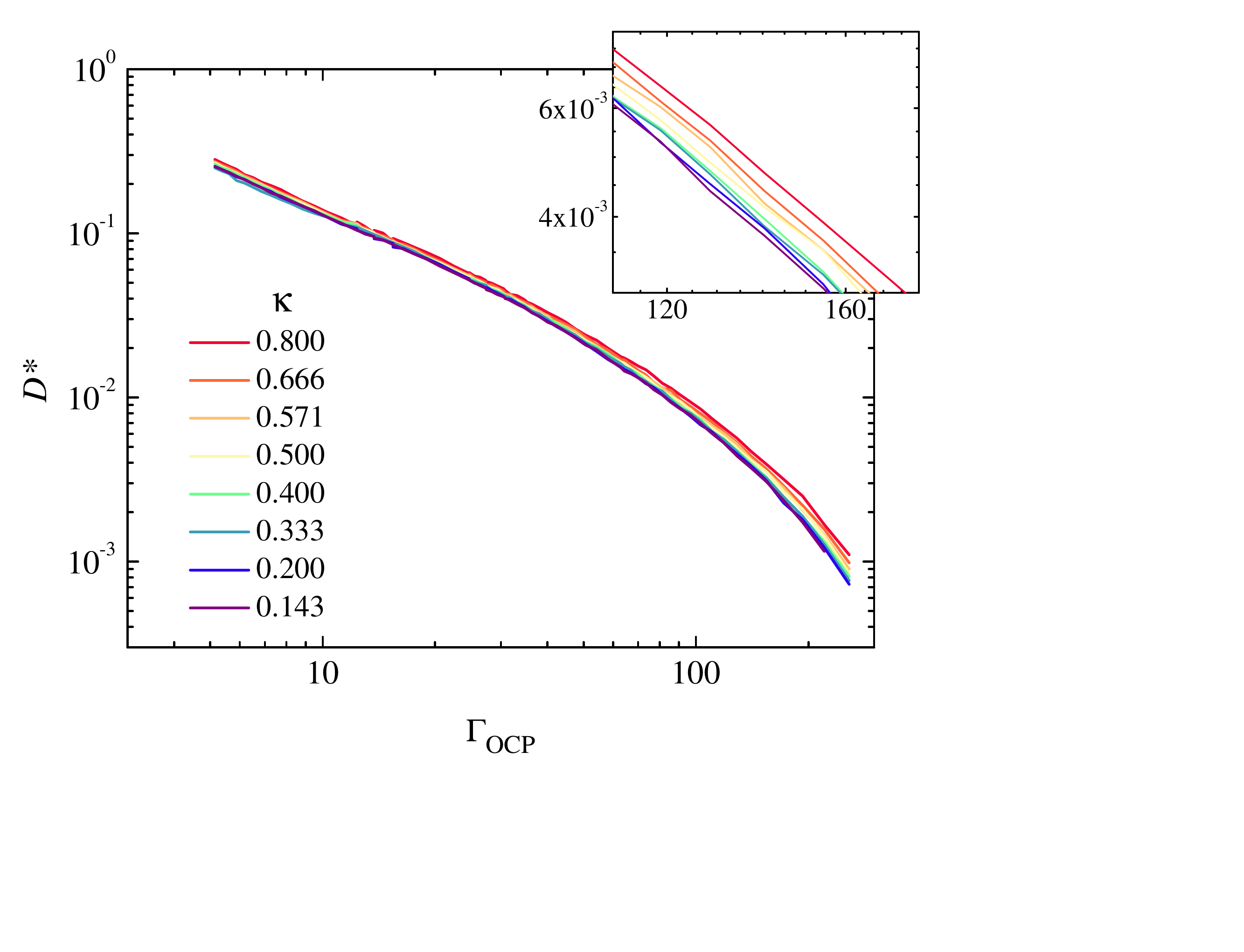}
\caption{Dimensionless diffusion coefficients in our MD simulations of OCPs {\new for various $\kappa=a/\lambda_e$ values} (data available in the supplemental materials). Inset shows behavior near crystallization. }	
\label{fig:MDOCP} 
\end{figure} 

\begin{figure*}
\centering  %left bottom top right)
\includegraphics[trim=0 120 180 0,clip,width=0.48\textwidth]{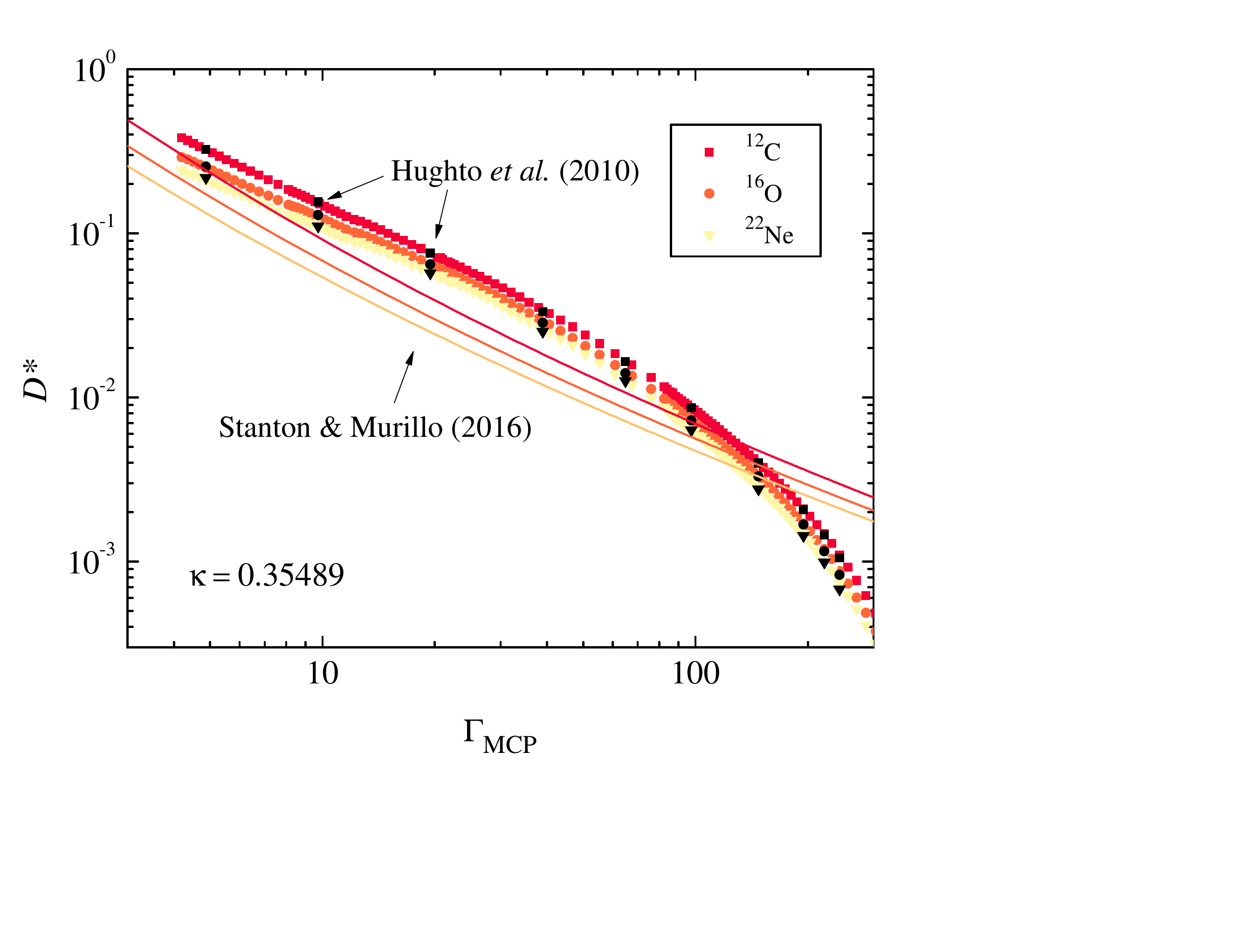}
\includegraphics[trim=0 120 180 0,clip,width=0.48\textwidth]{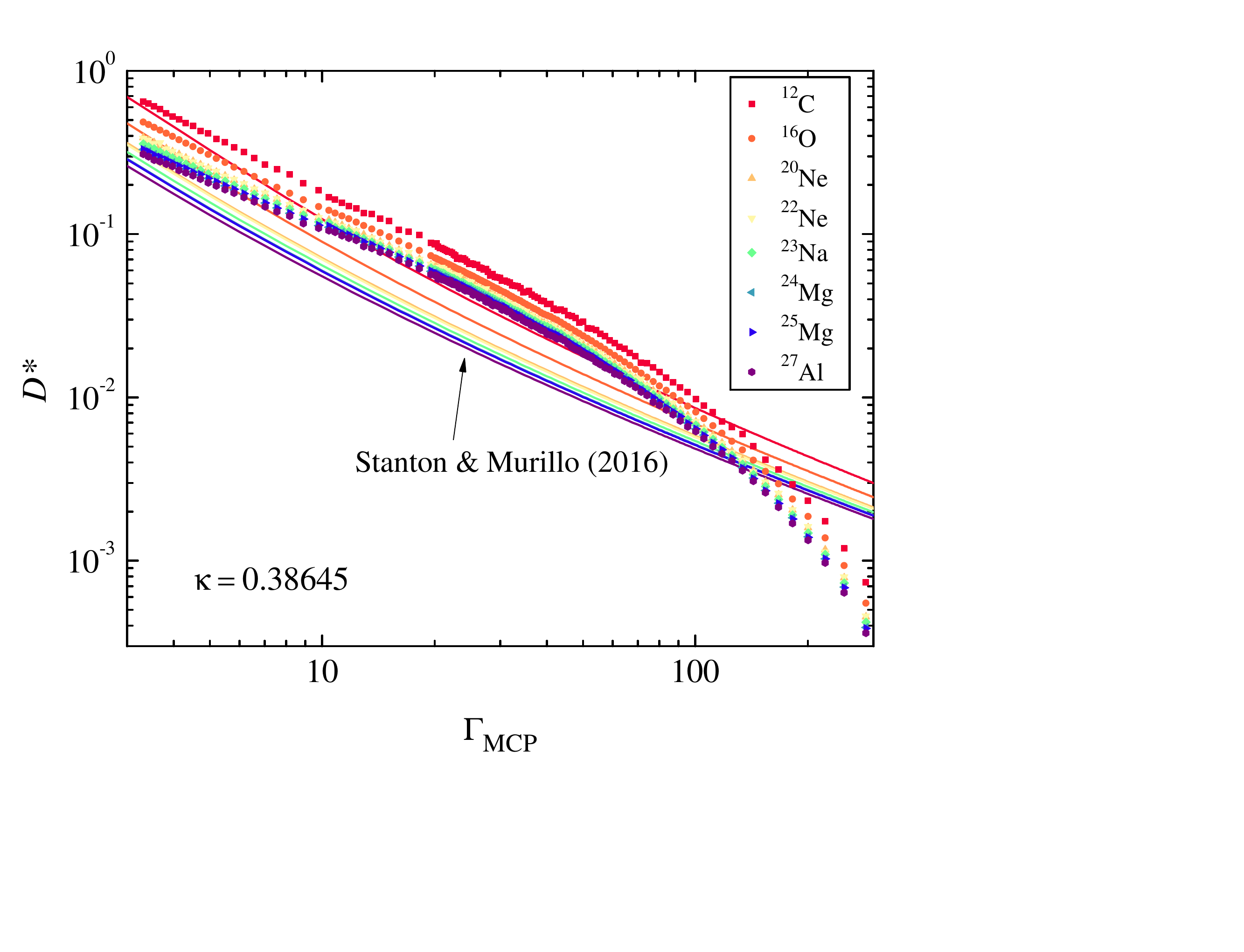}
\caption[]{  Diffusion coefficients for the C/O mixture of {\protect{\cite{Hughto2010}}} (left) and the O/Ne merger remnant mixture of {\protect{\cite{Schwab2021}}} (right) computed with MD (points) and the collision integral formalism (lines). We calculate $D^*(\Gamma_\mathrm{MCP})$ at roughly an order of magnitude more points in $\Gamma_{\mathrm{MCP}}$ than past work (data available in the supplemental materials). } 
\label{fig:MDmixtures}
\end{figure*}

Diffusion coefficients $D^*$ are calculated from the mean squared displacement (MSD), as in our past work \citep{caplan2021mnras}. The MSD is sensitive to the simulation time between the configurations. If this time is too small then we resolve only ballistic motion $\langle \mathbf x^2 \rangle \propto t^2$, while if this time is too large then particles become fully mixed over the period boundary. As such, for any given system we perform four simulations, each covering approximately a factor of 2-3 in $\Gamma$ with a configuration spacing chosen such that the mean displacement is several ion spacings but the maximum displacement is no greater than half a box length. While computationally efficient, we are limited at $\Gamma \lesssim 1$ by the long mean free path between collisions where the ballistic lengthscale becomes comparable to the box size. Likewise, $\Gamma \gtrsim 200$ the simulation becomes supercooled and simulating for long times becomes prohibitive because the configurations tend to spontaneously freeze. \footnote{In \cite{caplan2021mnras} we computed the MSD over an interval of $25 \omega_p^{-1}$ for all $\Gamma$. As $\Gamma \rightarrow 175$ we find that $D^*$ systematically underpredicts by about 10\% as the MSD is biased by oscillations on lattice sites rather than the lattice site hops. In this work we believe we have corrected for this and are in agreement with \cite{Daligault2012b}.}

In Fig. \ref{fig:MDOCP} we show diffusion coefficients for the OCP calculated at approximately 200 values of $\Gamma$ between $5 \lesssim \Gamma \lesssim 250$ for screening lengths $0 \lesssim \kappa \lesssim 1$, as in our past work. For $\Gamma < \Gamma_{\rm crit}$, there are effectively two regimes of diffusion. For $\Gamma \lesssim 1$ the plasma is weakly coupled and particles scatter in binary collisions, while for $\Gamma \gtrsim 1$ nuclei become trapped in `cages' of their neighbors and diffusion proceeds via thermally activated lattice site hops. The low $\Gamma$ limit returns the almost exact Chapman-Spitzer result, while at intermediate $\Gamma$ an Eyring model is appropriate. Together, this gives the power law behavior that bends downward at high $\Gamma$.

In Fig. \ref{fig:MDmixtures} we show the results of our MD of mixtures and compare to the diffusion coefficients calculated from Stanton \& Murillo. The first mixture is the same as \cite{Hughto2010}, with 49\% $^{12}$C, 49\% $^{16}$O, and 2\% $^{22}$Ne by number, and we include their simulation results for comparison. At a few $\Gamma$, \cite{Hughto2010} report multiple values for $D^*$ from simulations checking finite size effects which allows us to estimate their thermal noise is about 2\%, comparable to ours. Our MD generally agrees at this level at all $\Gamma$ checked, validating our scheme for computing $D^*$.
One point from \cite{Hughto2010}, at $\Gamma = 38.97$, is more than 5\% lower than our results here. This could just be a thermal fluctuation in their data, but it is hard to say given the spacing in their points. As noted by \cite{Bauer2020}, the fits from \cite{Stanton2016} tend to underpredict by up to a factor of 2 between $10 \lesssim \Gamma \lesssim 100$.

The second mixture is taken from \cite{Schwab2021} and is the result of carbon burning in a merger computed with MESA meant to be characteristic of a 1.35 $M_\odot$ merger remnant, as in \cite{Caiazzo2021}. By number abundance the mixture is 0.27\% $^{12}$C, 50.5\% $^{16}$O, 41.3\% $^{20}$Ne, and 3.8\% $^{24}$Mg, along with  0.18\% $^{22}$Ne, 2.4\% $^{23}$Na, 1.1\% $^{25}$Mg, and 0.46\% $^{27}$Al. The trace impurities have poor statistics when computing $D^*$ so we repeat our simulations 40 times with different random initial conditions and average the results, requiring nearly 4000 GPU hours.

\section{Model for Mixture Diffusion }\label{sec:model}

We start by presenting a fit that accurately describes diffusion in a liquid OCP, then provide a physically-motivated method for extending that fit to multi-component plasma mixtures.

%\subsection{OCP Fit}

Our model is adapted from \cite{Daligault2012b}, which consists of a piecewise defined $D^*(\Gamma)$ with different forms for low and high $\Gamma$. As noted in \cite{caplan2021mnras}, these two laws can easily be combined by taking the product of the Chapman-Spitzer result with an Eyring exponential. As such, we propose for the OCP  

\begin{equation}\label{eq:Dmodel}
    D_{\rm OCP}^*(\Gamma,\kappa) =  \sqrt{\frac \pi 3} \frac{ A(\kappa) \Gamma^{-5/2}}{ \ln \left( 1 + \frac{C(\kappa) }{ \sqrt{3}} \Gamma^{-3/2} \right) } e^{-B(\kappa) \Gamma}~,
\end{equation}

\noindent where $A(\kappa)$, $B(\kappa)$ and $C(\kappa)$ are given by

\begin{gather}
A (\kappa) = \sqrt{\frac \pi 3}(a_0  + a_1 \kappa^{a_2})~, \\
B (\kappa) = b_0 \exp( -b_1  \kappa^{b_2})~,  \\
C (\kappa) = c_0 + c_1 \rm \mathrm{erf}( c_2 \kappa^{c3} )~, 
\end{gather}

%\begin{subequations}
%\begin{equation}
%    A (\kappa) = \sqrt{\frac \pi 3}(a_0  + a_1 \kappa^{a_2})
%\end{equation}    
%\begin{equation}
%    B (\kappa) = b_0 \exp( -b_1  \kappa^{b_2}) 
%\end{equation}    
%\begin{equation}
%    C (\kappa) = c_0 + c_1 \rm \mathrm{erf}( c_2 \kappa^{c3} ) 
%\end{equation}
%\end{subequations}

\noindent parameterized by

\begin{gather}\notag
a_0 = 1.559 73, \quad a_1 = 1.109 41, \quad a_2 = 1.36909, \\
b_0 =  0.0070782,   \quad   b_1 = 0.80499
,   \quad   b_2  = 4.53523,    \\ \notag
c_0 = 2.20689,   \quad   c_1 = 1.351594, \\ \notag
\quad   c_2 = 1.57138,  \quad   c_3 = 3.34187. 
\end{gather}

The parameterizations of A and C are taken directly from \cite{Daligault2012b} are are valid for $0\leq \kappa \lesssim 4$, leaving us with just one free parameter B, which we fit to our new OCP MD. The parameterization of B was then generated from a least squares fit to Daligault's recommended form for $B(\kappa)$. We note that $B(\kappa)$ goes to zero more quickly in our model than in Daligault, as our $b_2$ is larger by a factor of about 2. As a consequence, the exponential suppression from the Eyring model has its onset at higher $\Gamma$. This is not such a major concern for WD cores, but future authors may prefer the original Daligault fits for $\kappa > 1$.

\begin{figure}
\centering  %left bottom top right)
\includegraphics[trim=0 120 180 0,clip,width=0.49\textwidth]{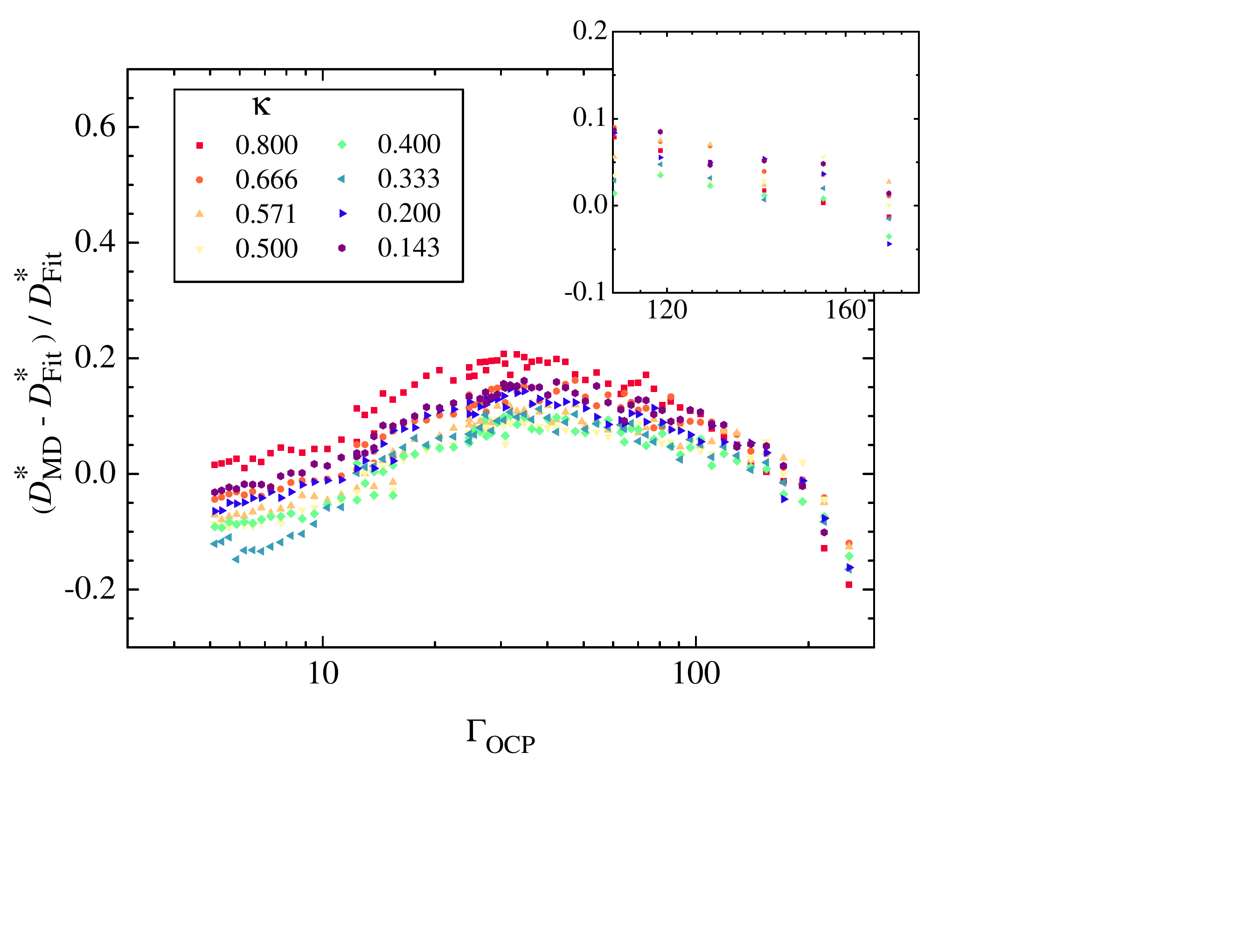}
\caption{\label{fig:OCPres} Normalized residuals between the OCP MD from Fig. \ref{fig:MDOCP} and the law from Eq. \ref{eq:Dmodel}. Inset shows the quality of the fit near crystallization.}	
\end{figure}

Fig. \ref{fig:OCPres} shows the residual of the fit to the OCP, and only in a few cases is the error larger than {\new 15\%}. For $\kappa = 0.333$ (typical of WD cores), the error is better than 10\% for $\Gamma > 10$. Near crystallization (inset) the fit works to within {\new 9\%} for all $\kappa$, and for $\kappa = 0.333$ it is good to 5\%. The fit appears to overpredict above $\Gamma_{\rm crit}$, but as discussed above this is more likely a result of the MD underpredicting due to supercooling of the fluid and the poor statistics of the lattice site hops. The overprediction of the fit between $10 \lesssim \Gamma \lesssim 100$ and underprediction at $\Gamma \lesssim 10$ are consistent with past work \citep{Daligault2012b,caplan2021mnras}. 

These results can be extended to plasma mixtures based on the following physical motivation. In the liquid regime of white dwarf interiors, \cite{Bildsten2001} suggested that even individual ions should behave as spheres with a radius determined by the ionic charge experiencing Brownian motion with a diffusion coefficient given by the Stokes-Einstein relation:
\begin{equation}
\label{eq:Stokes}
D = \frac{k_{\rm B} T}{4 \pi \eta R}~,
\end{equation}
where $\eta$ is the viscosity of the liquid and $R$ is the effective radius of an ion experiencing Stokes-Einstein drift. The MD results of \cite{Daligault2006} subsequently verified that this relation holds in a liquid OCP for $\Gamma \gtrsim 20$, with an effective ion radius on the order of the ion spacing in the OCP.

For a mixture in the liquid regime, each species of ion should therefore be expected to experience Stokes-Einstein drift. The viscosity is a global property that applies equally to any particle experiencing Brownian motion, so the only remaining parameter setting the diffusion coefficient of a species according to Eqn~\eqref{eq:Stokes} is its effective radius. Since particle interactions are mediated by Coulomb forces, the effective radius will depend primarily on a particle's charge relative to the background average, with particles of higher charge having a larger effective radius due to stronger Coulomb interactions, as suggested by \cite{Bauer2020}.

We therefore propose the following simple generalization of our diffusion law for OCPs to MCPs that may be readily implemented in stellar evolution codes. The diffusion coefficient of a species $j$ with charge $Z_j$ can be described by treating the background liquid as having the effective viscosity of an equivalent OCP with average charge $\langle Z \rangle$. The resulting diffusion coefficient can then be written in terms of that for an equivalent OCP, but rescaled according to charge to reflect the dependence of the effective radius on $Z_j$:
\begin{equation}\label{eq:Dmodel2} 
    D_j^*(\Gamma_{\rm MCP},\kappa) = D_{\rm OCP}^*(\Gamma_{\rm MCP},\kappa) \left( \frac{Z_j}{ \langle Z \rangle} \right)^{-0.6}~,
\end{equation}
which is similar to the scaling suggested in \cite{Hughto2010}. The normalization for the equivalent OCP uses the mixture averaged quantities for $\omega_p a^2$ . The scaling with charge to the power of -0.6 is chosen empirically to provide a good match to the MD, and we do not currently have a physical explanation for how we should expect effective particle radius (and therefore diffusion coefficient) to scale with charge {\it a priori}. This parameter was not fit in this work but future authors could. Our MD suggests that this scaling results in at worst 10\% spread for charge ratios typical of WDs, though future work may like to verify this behavior out to large charge ratios, for example in mixtures including significant He or Fe.

\begin{figure*}
\centering  %left bottom top right)
\includegraphics[trim=0 120 180 0,clip,width=0.48\textwidth]{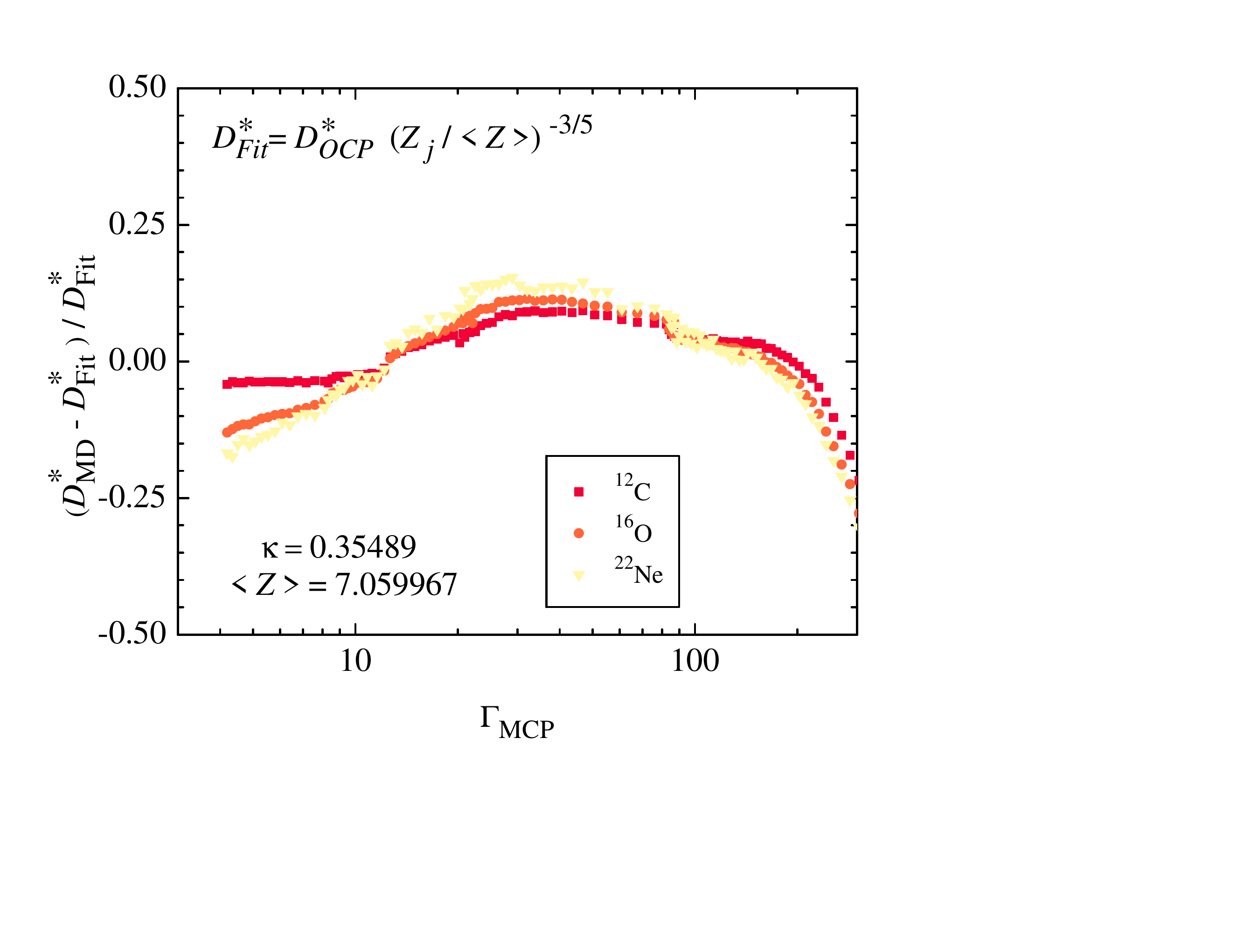}
\includegraphics[trim=0 120 180 0,clip,width=0.48\textwidth]{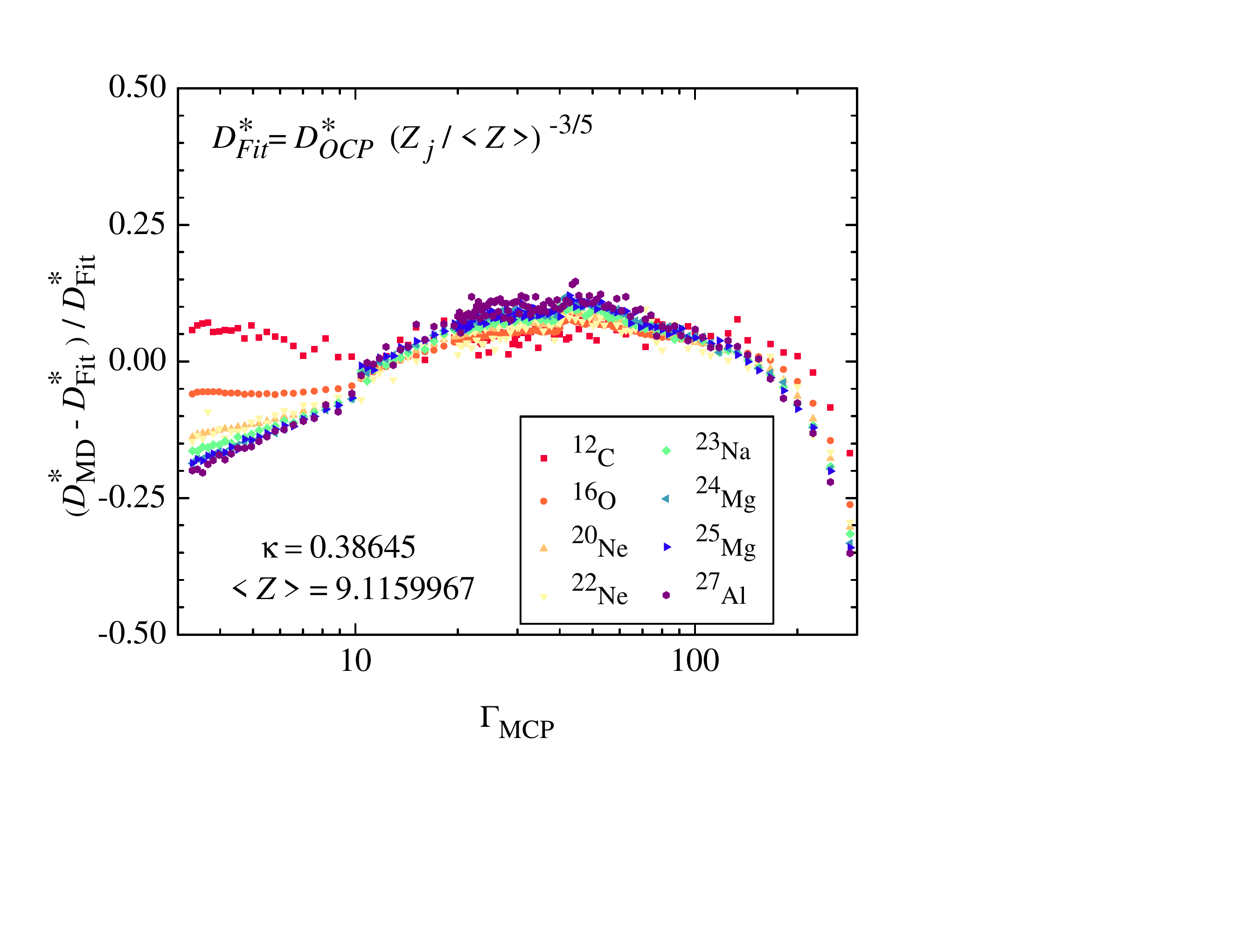}
\caption{\label{fig:MDres} Normalized residuals between the mixture MD in Fig. \ref{fig:MDmixtures} and Eq. \ref{eq:Dmodel2}.}
\end{figure*}

In Fig. \ref{fig:MDres} we show the quality of the fit for the two mixtures. Our proposed fit is good to within {\new 15\%} for all species over the entire strongly coupled regime $10 \lesssim \Gamma \lesssim 200$ where the Stokes-Einstein relation is expected to hold. For $5 \lesssim \Gamma \lesssim 10$ the fit tends to separate, but remains better than 25\%. This may not be so important, as at $\Gamma \lesssim 1$ the behavior becomes that of a collisional ideal gas and fairly exact solutions are known. {\new Near crystallization ($70 \lesssim \Gamma \lesssim 200$)}, the model is again good to 10\% for all species in both mixtures, which is most relevant for sedimentary heating and the separation processes that occur near or at crystallization. We also note that the O/Ne merger remnant mixture contains two isotopes each of Ne and Mg, allowing us to check how neutron excesses may impact diffusion of trace impurities in mixtures. When comparing $D^*$ of $^{20}$Ne to $^{22}$Ne and $^{24}$Mg to $^{25}$Mg we find that both pairs of isotopes show no difference between the symmetric nucleus and the neutron-excess nucleus within the level of our thermal noise. Therefore, the mixture model of Eq. \ref{eq:Dmodel2} is suitable for modeling the behavior of neutron rich nuclei (\ie\ $Z/A < 0.5$) in WDs despite the model only being dependent on charge and not explicitly accounting for the mass ratios.

\section{Conclusion}\label{sec:dis}

In this work we have developed a new model for diffusion in WDs that is capable of accurately computing diffusion coefficients for multi-component mixtures to {\new 15\%} accuracy. We expect this model will be most useful for screening lengths $0 \lesssim \kappa \lesssim 1$ and coupling $10 \lesssim \Gamma_{\rm MCP} \lesssim 200$, but it is possible that this model will work well beyond these regions given the theoretical basis for the functional form of the model and the past work by \cite{Daligault2012b} developing the parametrization. 
In addition this model's high accuracy in the moderate to strongly coupled regime it is also almost trivial to implement in stellar evolution codes, requiring only one additional computational operation per species in the mixture. Our equations in sec. \ref{sec:model} are efficient, especially in complicated mixtures with many trace components. This eliminates the need for computing large numbers of binary-collision integrals, which can become computationally expensive in realistic mixtures with many elements.

While the current model is suitable for WD interiors, future work should extend the parameter space checked by MD to validate this model at larger screening lengths, charge ratios, and even mass ratios. In our O/Ne mixture, no species differed from the average mixture charge by more than a factor of 1.4. Proposals that iron precipitation may be relevant in C/O WDs motivates the study of charge ratios as large as 4 \citep{Caplan2021iron}. 

%Astrophysically, this work may be most immediately relevant to late time heating from the sedimentation of neutron rich nuclei and separation processes which can drive precipitation or distillation in the core. Comparing to $D^*$ from \cite{Stanton2016}, we see that the curves intersect near $\Gamma \approx 100$, so the heating and luminosity excess prior to the onset of crystallization won't be too different from what past work has calculated \citep{Bauer2020,Schwab2021}, but the larger $D^*$ atintermediate Gamma could result in faster transport toward the core at slightly earlier times. This could change abundances which impact precipitation/distillation processes, and also widen the mass range of mergers susceptible to `delayed' supernova by transporting greater amounts of neutron rich nuclei to the core thus driving a greater increase in the core electron Fermi energy. 

\section*{Acknowledgements}
We thank Earl Bellinger and Lars Bildsten for helpful comments on a draft of this manuscript.
This research was supported in part by supercomputing resources provided by Illinois State University.
The authors acknowledge the Indiana University Pervasive Technology Institute for providing supercomputing resources that have contributed to the research results reported within this paper. This research was supported in part by Lilly Endowment, Inc., through its support for the Indiana University Pervasive Technology Institute. 

%%%%%%%%%%%%%%%%%%%%%%%%%%%%%%%%%%%%%%%%%%%%%%%%%%
\section*{Data Availability}

The data underlying this article are available in the article and in its online supplementary material.



\bibliographystyle{mnras}
%\bibliography{apsbib} % if your bibtex file is called example.bib

\providecommand{\noopsort}[1]{}\providecommand{\singleletter}[1]{#1}%

%\begin{thebibliography}{99}
%\end{thebibliography}

%%%%%%%%%%%%%%%%%%%%%%%%%%%%%%%%%%%%%%%%%%%%%%%%%%

%%%%%%%%%%%%%%%%% APPENDICES %%%%%%%%%%%%%%%%%%%%%

%\appendix


%If you want to present additional material which would interrupt the flow of the main paper, it can be placed in an Appendix which appears after the list of references.

%%%%%%%%%%%%%%%%%%%%%%%%%%%%%%%%%%%%%%%%%%%%%%%%%%

% Don't change these lines
\bsp	% typesetting comment
\label{lastpage}
\end{document}